\documentclass[journal]{IEEEtran}
\IEEEoverridecommandlockouts
\usepackage{pgfplots}
\pgfplotsset{compat=1.18}
\usepackage{xcolor}
\usepackage{pifont}
 \usepackage{url}
 \usepackage[most]{tcolorbox}

\newtcolorbox{rqbox1}{
    colback=gray!8,
    colframe=black,
    boxrule=0.8pt,
    arc=2pt,
    left=6pt,
    right=6pt,
    top=6pt,
    bottom=6pt,
    title=System Prompt,
    fonttitle=\bfseries
}
\definecolor{GOne}{RGB}{33,113,181}     
\definecolor{GTwo}{RGB}{35,139,69}      
\definecolor{GThree}{RGB}{146,36,40}    
\definecolor{GFour}{RGB}{117,107,177}   
\definecolor{GFive}{RGB}{217,95,2}      
\definecolor{EdgeColor}{RGB}{0,90,120}      
\definecolor{EdgeColorLight}{RGB}{0,130,160}
\usepackage{booktabs}
\usepackage{cite}
\usepackage{amsmath,amssymb,amsfonts}
\usepackage{algorithmic}
\usepackage{graphicx}
\usepackage{textcomp}
\usepackage{xcolor}
\def\BibTeX{{\rm B\kern-.05em{\sc i\kern-.025em b}\kern-.08em
    T\kern-.1667em\lower.7ex\hbox{E}\kern-.125emX}}
\begin{document}

\title{How Small Can 6G Reason? Scaling Tiny-to-Small Language Models for AI-Native Networks
}
\author{
\IEEEauthorblockN{
Mohamed~Amine~Ferrag\IEEEauthorrefmark{1}\IEEEauthorrefmark{3},
Abderrahmane~Lakas\IEEEauthorrefmark{1},
Merouane~Debbah\IEEEauthorrefmark{2}
}
\\
\IEEEauthorblockA{\IEEEauthorrefmark{1}
Department of Computer and Network Engineering,  
United Arab Emirates University, UAE
} \\
\IEEEauthorblockA{\IEEEauthorrefmark{2}
Research Institute for Digital Future, Khalifa University, UAE
}\\
\IEEEauthorblockA{\IEEEauthorrefmark{3}
Corresponding author: \texttt{mohamed.ferrag@uaeu.ac.ae}
}
}

\maketitle

\begin{abstract}
Emerging 6G visions, reflected in ongoing standardization efforts within 3GPP, IETF, ETSI, ITU-T, and the O-RAN Alliance, increasingly characterize networks as AI-native systems in which high-level semantic reasoning layers operate above standardized control and data-plane functions. Although frontier-scale large language models (LLMs) such as Qwen2.5-7B and Olmo-3-7B demonstrate strong reasoning capability, their computational footprint limits deployment in latency-sensitive, edge-native infrastructures. This paper presents a systematic empirical study of the scaling behavior and deployment efficiency of compact language models for network-level semantic reasoning in AI-native 6G systems. Using 6G-Bench—a standardization-aligned benchmark comprising 30 decision-making tasks across five capability domains—we evaluate models ranging from 135M (SmolLM2-135M) to 7B parameters (Qwen2.5-7B), including mid-scale architectures such as Llama-3.2-1B, Granite-1B, and Qwen2.5-3B. Deterministic accuracy (pass@1) increases from 0.224 at 135M to 0.707 at 7B, but scaling gains are highly non-uniform. {\color{black} A pronounced stability transition occurs in the 1--1.5B range, where accuracy rises from 0.373 (Llama-3.2-1B) to 0.531 (Qwen2.5-1.5B) and the instability gap $\Delta_5$ contracts from 0.356 to 0.138. Across the full evaluated spectrum (135M--7B), $\Delta_5$ further decreases from 0.365 to 0.031, indicating increasing alignment between stochastic exploration and deterministic inference. } Through single-query inference profiling and an Edge Score metric that normalizes accuracy by latency and memory footprint, we show that semantic reliability per unit edge resource does not scale monotonically with parameter count. Instead, mid-scale models (approximately 1.5--3B) achieve the most favorable balance between deterministic stability and computational efficiency, providing deployment-relevant guidance for AI-native 6G architectures. All scripts and results are publicly available at \url{https://github.com/maferrag/6G-Bench}.
\end{abstract}


\section{Introduction}

The evolution toward AI-native 6G networks represents a structural shift in how intelligence is embedded into communication systems, elevating artificial intelligence from a localized optimization tool to a native architectural layer spanning radio access networks (RAN), core networks, and distributed edge–cloud infrastructures. Within 3GPP \cite{3gpp_tr_22_870_2025}, service-based architectures, intent-driven management, network exposure functions (NEF), slicing, and digital twins reflect the emergence of semantic control layers above standardized protocol stacks, while the O-RAN Alliance \cite{ORAN-nGRG-GenAI-6G-2025} promotes AI-driven closed-loop control through near-real-time and non-real-time RIC controllers. ETSI MEC \cite{etsi_gs_mec_003_v3_1_1_2022} formalizes edge-native orchestration, and IETF efforts \cite{ietf_hw_ai_agent_6g_00} on intent-based networking, autonomous systems, and zero-trust security, together with ITU-T IMT-2030 visions \cite{hossain20256g}, target sub-millisecond latency ($<$1 ms), peak rates approaching 0.1–1 Tbps, reliability beyond 99.99999\%, and pervasive AI integration. At the physical and system levels, disruptive technologies—including sub-THz/THz operation, extremely large antenna arrays (ELAA), cell-free massive MIMO \cite{lu2025energy}, reconfigurable intelligent surfaces (RIS) enabling programmable propagation \cite{banerjee20266g}, integrated sensing and communication, semantic communication, federated learning, digital twins, and quantum-safe or quantum-assisted networking \cite{sharif2026resource,turnip2025towards}—dramatically increase network dimensionality and controllability. Coordinating RIS phase shifts, THz beam alignment, slice arbitration, digital twin synchronization, and quantum-secure trust management requires multi-step reasoning across radio, compute, and policy layers, motivating the exploration of LLMs and AI agents as semantic orchestration engines \cite{gogineni2025llms,ferrag2026alpha}. However, frontier-scale models (70B–500B parameters) demand multi-accelerator clusters and hundreds of gigabytes of memory, making them impractical for latency- and power-constrained RAN or MEC deployments \cite{liu2024mobilellm,wu2024netllm}.

Scaling laws motivate a fundamental question for AI-native 6G architecture design:
\emph{How does network-level semantic reasoning capability scale in small and tiny language models under 6G constraints?}
Unlike conventional QA benchmarks, 6G decision-making requires structured reasoning across heterogeneous abstractions, including intent feasibility under dynamic network state, slice selection and resource arbitration, RIS and massive MIMO configuration, THz beam management, SLA violation forecasting, trust-aware compute exposure, quantum-secure authentication, inter-operator federation, distributed AI coordination, and ISAC-driven control. These capabilities align with architectural trajectories articulated in 3GPP service-based systems, IETF intent networking drafts \cite{ietf_hw_ai_agent_6g_00}, ETSI MEC orchestration frameworks \cite{etsi_gs_mec_003_v3_1_1_2022}, O-RAN intelligent control loops \cite{ORAN-nGRG-GenAI-6G-2025}, and ITU IMT-2030 visions \cite{hossain20256g}. Determining whether stable deterministic reasoning emerges at 0.5B, 1B, or 3B parameters—and quantifying the marginal gains beyond this regime—therefore becomes not only a scaling-law investigation, but a foundational architectural problem for deploying AI agents across hierarchical 6G, edge, and quantum-enhanced network tiers.

In this work, we conduct a systematic scaling analysis of small and tiny language models (135M–7B parameters) using 6G-Bench \cite{ferrag20266g}, a standardization-aligned evaluation framework comprising 30 decision-making tasks organized into five capability domains derived from 3GPP, IETF, ETSI, ITU-T, and O-RAN architectural directions. Rather than proposing a new benchmark, we leverage this controlled evaluation environment to characterize empirical scaling regimes governing deterministic semantic reasoning in AI-native 6G contexts. Our study explicitly targets the deployment-relevant regime below 10B parameters—far smaller than frontier LLMs (70B–500B)—and focuses on safety-critical, policy-constrained network reasoning under zero-shot, deterministic decoding. We quantify how pass@1 accuracy evolves from 0.224 at 135M parameters to 0.707 at 7B, identify a pronounced stability transition in the 1–1.5B range, and analyze how reasoning instability ($\Delta$5) contracts from 0.365 to 0.031 as scale increases. This enables principled characterization of when compact LLM-based AI agents become sufficiently reliable for hierarchical 6G control-plane deployment.

\begin{table*}[t]
\centering
\caption{Comparison of scaling studies and their scope relative to AI-native 6G semantic reasoning.}
\label{tab:related_comparison}
\scriptsize
\begin{tabular}{lcccc}
\toprule
Aspect & Kaplan et al.  \cite{kaplan2020scaling} & Hoffmann et al. \cite{hoffmann2022training} & Wei et al. \cite{wei2022emergent} & This Work \\
\midrule
Primary Objective & LM loss scaling & Compute-optimal LM & Emergent abilities & 6G semantic reasoning \\
Evaluation Metric & Cross-entropy & Cross-entropy & Downstream accuracy & pass@1, pass@k, $\Delta_k$ \\
Scale Range & $10^3$–$10^9$ & 70M–280B & 10B–500B & 135M–7B \\
AI-Native 6G Task Alignment & \ding{55} & \ding{55} & \ding{55} & \ding{51} \\
Deterministic Stability Analysis & \ding{55} & \ding{55} & \ding{55} & \ding{51} \\
Domain-Specific Scaling Elasticity & \ding{55} & \ding{55} & \ding{55} & \ding{51} \\
Edge Deployment Focus & \ding{55} & \ding{55} & \ding{55} & \ding{51} \\
\bottomrule
\end{tabular}
\end{table*}

The remainder of this paper is organized as follows. 
Section \ref{sec:related} reviews classical scaling laws for large language models and discusses their limitations in the context of safety-critical, AI-native 6G network reasoning.  Section \ref{sec:methodo} presents the methodology, including the 6G-Bench task taxonomy, the evaluated model spectrum (135M–7B parameters), the experimental protocol, and the formal definitions of deterministic and stochastic stability metrics. 
Section \ref{sec:Scaling} provides the empirical scaling analysis, covering global performance trends, instability-gap contraction, domain-specific scaling elasticity, and deployment-tier implications for hierarchical 6G architectures. Section \ref{sec:conclu} concludes the paper and outlines directions for future research on compact LLM-based AI agents for edge-native, quantum-secure 6G systems.

\section{Related Work}
\label{sec:related}

Recent research on large language models has established strong empirical regularities governing performance as a function of model size, training data, and compute \cite{bommasani2021opportunities,bubeck2023paper,srivastava2023beyond,wang2022self}. Table~\ref{tab:related_comparison} presents a comparison of prior language model scaling studies and highlights their differences relative to AI-native 6G semantic reasoning requirements.

\subsection{Scaling Laws for Large Language Models}

Kaplan et al. \cite{kaplan2020scaling} established that autoregressive Transformer performance follows stable power-law relationships with respect to model size, dataset size, and compute, showing that test loss scales as $L(N) \propto N^{-\alpha_N}$ with $\alpha_N \approx 0.076$, and similarly with data and compute under optimal allocation. They derived a unified formulation $L(N,D)$ capturing overfitting behavior, demonstrating that data requirements grow sublinearly with model size ($D \propto N^{0.74}$) and that compute-optimal scaling strongly favors increasing model size ($N \propto C_{\min}^{0.73}$), with minimal growth in training steps. Architectural variations were found to have a comparatively minor impact relative to scale. While this framework provides a predictive theory of perplexity scaling across orders of magnitude, it focuses exclusively on next-token prediction and does not address deterministic reliability, policy-consistent reasoning, or safety-critical semantic control tasks central to AI-native 6G systems—leaving open how such scaling laws translate to structured, network-level reasoning in compact models.

\begin{figure*}[t]
    \centering
    \includegraphics[width=0.9\textwidth]{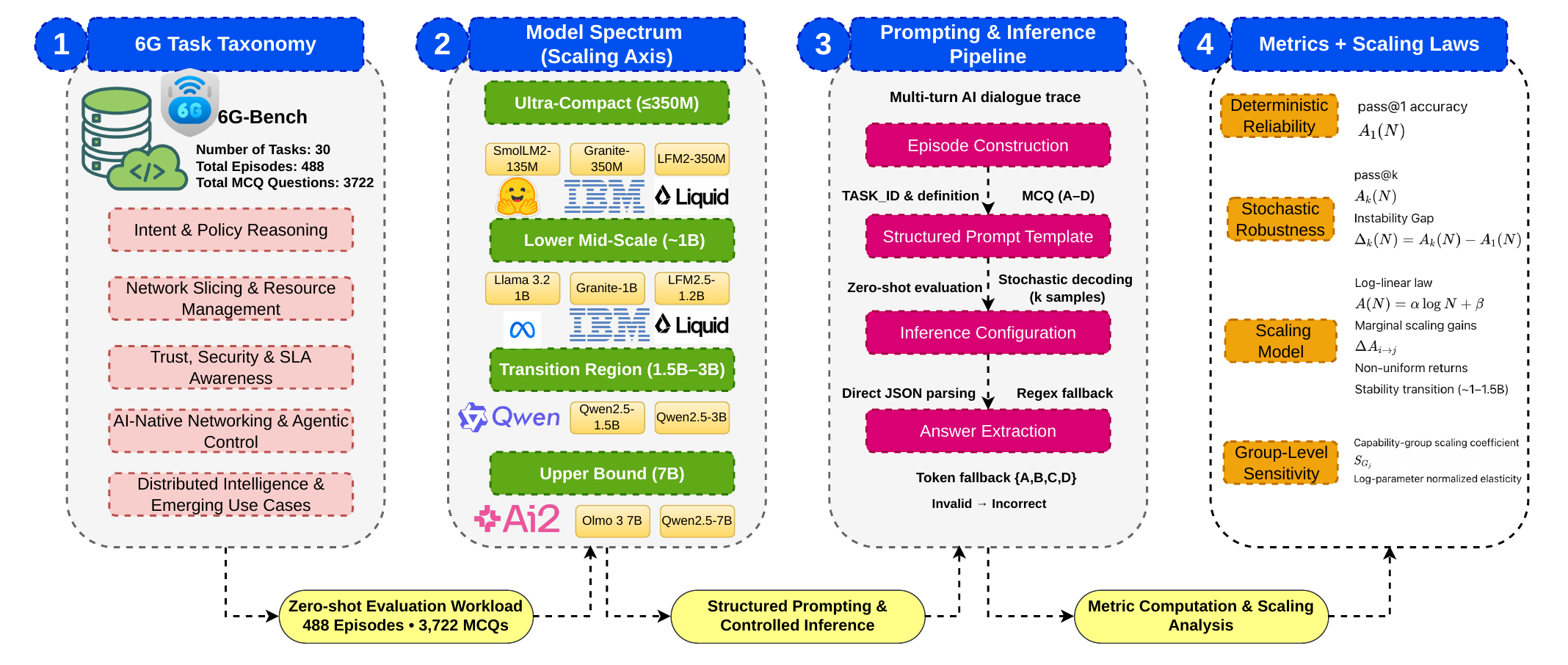}
    \caption{End-to-End Methodological Framework for Evaluating Parameter Scaling and Semantic Reasoning Stability in 6G-Bench.}
    \label{fig:fig1}
\end{figure*}

\subsection{Compute-Optimal Scaling and Data-Parameter Trade-offs}

Hoffmann et al. \cite{hoffmann2022training} revisited compute-optimal scaling by modeling $L(N,D)$ under fixed FLOPs budgets using over 400 models (70M–16B parameters, 5B–500B tokens). Contrary to earlier predictions that parameters should grow faster than data, they showed that optimal performance requires near-equal scaling of parameters and tokens, with $N_{\text{opt}} \propto C^{0.5}$ and $D_{\text{opt}} \propto C^{0.5}$. Their results indicated that many large models were undertrained, a hypothesis validated by Chinchilla (70B parameters, 1.4T tokens), which outperformed the 280B-parameter Gopher model on downstream benchmarks, including a 67.6\% MMLU average, while using the same compute budget and enabling lower inference cost. Although this work refined scaling-law prescriptions for compute-efficient training, it remained centered on language modeling loss and standard NLP evaluations.

\subsection{Emergent Abilities and Phase Transitions in Scaling}

Wei et al. \cite{wei2022emergent} analyzed \emph{emergent abilities}, identifying qualitative capability transitions that arise only beyond specific scale thresholds, deviating from smooth extrapolation trends. Across benchmarks, they observed phase-transition-like behavior when scaling training compute or parameter count: tasks such as modular arithmetic and MMLU showed near-random performance below $\sim10^{22}$ FLOPs ($\approx$1 10B–13B parameters), followed by sharp gains in the $10^{23}$ FLOP regime ($\approx$1 50B–280B parameters), where models such as Chinchilla \cite{hoffmann2022training} and Gopher \cite{rae2021scaling} achieved substantial improvements. Similar emergence patterns were reported for TruthfulQA, chain-of-thought prompting, instruction tuning, scratchpad reasoning, calibration, and multilingual tasks. These findings suggest that large-scale capability shifts can occur abruptly at high compute thresholds, though their implications for structured, safety-critical reasoning in constrained deployment regimes remain unclear.

\subsection{Limitations of Prior Scaling Studies for AI-Native 6G}

Despite the substantial progress in understanding language model scaling, prior studies remain fundamentally centered on open-domain NLP objectives and token-level performance metrics. Kaplan et al.~\cite{kaplan2020scaling} and Hoffmann et al.~\cite{hoffmann2022training} derive predictive power-law relationships for cross-entropy loss under varying parameter, data, and compute regimes, while Wei et al.~\cite{wei2022emergent} analyze qualitative capability transitions across downstream reasoning benchmarks. However, these works evaluate perplexity, few-shot accuracy, or prompting-based task performance rather than deterministic decision reliability under structured, safety-critical constraints. AI-native 6G networks introduce distinct requirements: semantic reasoning must operate within latency constraints, adhere to policy enforcement rules, operate under partial observability, satisfy SLA constraints, and coordinate across heterogeneous domains. Tasks such as intent feasibility assessment, trust-aware compute exposure, slice arbitration, and distributed intelligence orchestration are not equivalent to standard NLP benchmarks and demand consistent single-shot correctness and stability under deterministic decoding. Furthermore, existing scaling analyses primarily focus on large-scale regimes (tens to hundreds of billions of parameters), leaving the sub-billion- to low-billion-parameter range—most relevant for edge-native deployment—largely uncharacterized in terms of semantic control reliability. Consequently, there is currently no empirical framework that examines scaling regimes, stability thresholds, and domain-specific scaling elasticity for structured network-level reasoning in AI-native 6G systems, motivating the dedicated small-model scaling analysis conducted in this work.

\section{Methodology}
\label{sec:methodo}

This section describes the experimental framework used to evaluate the scaling behavior of small and mid-scale language models for network-level semantic reasoning in AI-native 6G systems. We first present the task taxonomy and capability grouping underlying 6G-Bench, followed by the evaluated model spectrum spanning 135M to 7B parameters. We then detail the structured prompting protocol, the construction of the dataset, the inference configuration, and the reproducibility controls used for zero-shot evaluation across 488 episodes and 3{,}722 multiple-choice questions. Finally, we formalize deterministic and stochastic evaluation metrics, introduce the instability gap, and define log-linear and group-level scaling models used to characterize parameter-dependent reasoning regimes. Fig. \ref{fig:fig1} presents the complete experimental framework for evaluating scaling-dependent semantic reasoning in AI-native 6G systems.

\subsection{Task Taxonomy and Capability Groups}

The 6G-Bench benchmark \cite{ferrag20266g} defines 30 decision-making tasks (T1–T30) organized into five capability-aligned categories derived from ongoing standardization efforts in 3GPP, IETF, ETSI, ITU-T, and the O-RAN Alliance. The five capability groups are:

\begin{itemize}
    \item \textbf{Intent \& Policy Reasoning}: semantic consistency under evolving network state and policy constraints.
    \item \textbf{Network Slicing \& Resource Management}: slice selection, compute placement, SLA forecasting, and resource arbitration.
    \item \textbf{Trust, Security \& SLA Awareness}: trust-aware exposure, authorization reasoning, and automated security response.
    \item \textbf{AI-Native Networking \& Agentic Control}: inter-agent coordination, federation, lifecycle management, and task offloading.
    \item \textbf{Distributed Intelligence \& Emerging Use Cases}: federated learning orchestration, Integrated Sensing and Communication (ISAC), digital twins, and public-safety coordination.
\end{itemize}

\begin{table*}[t]
\centering
\caption{Evaluated model spectrum across parameter scale, organization, architecture, and alignment pipeline. All models were downloaded from Hugging Face and evaluated locally under a unified inference protocol.}
\scriptsize
\label{tab:model_spectrum}
\begin{tabular}{lccccccc}
\hline
Model & Organization & Params (B) & Architecture & Alignment & Context & Target Deployment & Ref. \\
\hline
SmolLM2-135M & Hugging Face & 0.135 & Dense Transformer & SFT + DPO & 8K & Ultra-edge & \cite{allal2025smollm2} \\
Granite-4.0-350M & IBM & 0.35 & Dense Transformer & SFT + RL + Merge & 8K & On-device & \cite{granite2025} \\
LFM2-350M & Liquid AI & 0.35 & Hybrid (Conv + Attn) & Post-train & 8K & Edge CPU/NPU & \cite{liquidai2025lfm2} \\
\hline
Llama 3.2 1B & Meta & 1.0 & Dense Transformer & Instruct SFT & 8K & Edge/Control & \cite{grattafiori2024llama} \\
Granite-4.0-H-1B & IBM & 1.0 & Dense Transformer & SFT + RL & 8K & Domain-adaptive & \cite{granite2025} \\
LFM2.5-1.2B & Liquid AI & 1.2 & Hybrid (LIV + GQA) & Ext. Pretrain + RL & 32K & Edge long-context & \cite{liquidai2025lfm2} \\
\hline
Qwen2.5-1.5B & Alibaba & 1.5 & Dense (RoPE, SwiGLU) & Pretrain + Post & 32K & Control-plane & \cite{qwen25,qwen2} \\
Qwen2.5-3B & Alibaba & 3.0 & Dense (RoPE, SwiGLU) & Pretrain + Post & 32K & Operator domain & \cite{qwen25,qwen2} \\
Olmo 3 7B & AI2 & 7.0 & Dense Transformer & Instruct SFT & 8K & Central control & \cite{olmo2025olmo} \\
Qwen2.5-7B & Alibaba & 7.0 & Dense (RoPE, SwiGLU) & Pretrain + Post & 32K & Central orchestration & \cite{qwen25,qwen2} \\
\hline
\end{tabular}
\\
Abbreviations: SFT = Supervised Fine-Tuning; 
RL = Reinforcement Learning; 
DPO = Direct Preference Optimization; 
Merge = Model Merging; 
Conv = Convolution; 
Attn = Attention; 
LIV= linear input-varying operators; 
GQA = Grouped-Query Attention; 
RoPE = Rotary Positional Embedding; 
SwiGLU = Swish-Gated Linear Unit; 
Pretrain = Pre-training; 
Post = Post-training alignment.
\end{table*}

\subsubsection{Task Formatting, Answer Extraction, and Grading Protocol}

Each evaluation instance consists of a structured episode--question pair. Episodes are stored as JSON objects containing (i) an initial system state (environment, airspace, UAV configuration, and policy constraints), and (ii) a multi-turn dialogue trace including intents, actions, network telemetry, and tool responses. For each episode, one or more multiple-choice questions (MCQs) are defined, each associated with a task identifier (T1--T30), four answer options (A--D), a ground-truth label, and metadata including the source turn and difficulty.

To ensure controlled reasoning and reproducible grading, the model is prompted using a structured template comprising:
\begin{itemize}
    \item Target task identifier and task definition,
    \item A summarized episode context,
    \item The MCQ question text,
    \item Four labeled answer options (A--D), and
    \item A strict output instruction requiring valid JSON of the form \texttt{\{"answer": "A"\}}.
\end{itemize}

The episode summary is programmatically generated to include:
\begin{itemize}
    \item Initial environmental, airspace, UAV, and policy context,
    \item Mission success indicator,
    \item A truncated dialogue trace (up to 12 turns), and
    \item Aggregated network telemetry per turn (slice, latency, jitter, loss, throughput, edge load).
\end{itemize}

The model must return a single JSON object. Output parsing follows a deterministic extraction pipeline: (i) attempt direct JSON parsing; (ii) if parsing fails, apply regular-expression-based extraction of the \texttt{"answer"} field; and (iii) as a fallback, extract a standalone token in \{\texttt{A}, \texttt{B}, \texttt{C}, \texttt{D}\}. A prediction is considered valid only if it resolves to one of the four options; all other outputs are marked as incorrect.

\subsection{Model Spectrum and Experimental Protocol}

To systematically characterize the scalability of semantic reasoning under AI-native 6G constraints, we evaluate a diverse set of compact and mid-scale language models ranging from 135M to 7B parameters. The selected models span multiple architectural paradigms, training regimes, and deployment targets, enabling controlled analysis of parameter scaling effects and architectural efficiency. Table~\ref{tab:model_spectrum} summarizes the evaluated model spectrum, detailing parameter scale, organizational origin, architectural design, alignment strategy, and deployment orientation.

\subsubsection{Model Selection Rationale}

The model spectrum was constructed to satisfy four methodological objectives: 
(i) granular coverage of the sub-1B regime, where edge-native deployment is most realistic; 
(ii) dense sampling of the 1–1.5B transition region, hypothesized to represent a stability boundary; 
(iii) inclusion of mid-scale (3B) and upper-bound (7B) models to identify diminishing returns; and 
(iv) architectural heterogeneity, including conventional dense Transformer models and hybrid convolution-attention designs optimized for edge inference. All evaluated models are instruction-tuned variants, reflecting the intent-driven and policy-conditioned nature of 6G semantic reasoning tasks.

\subsubsection{Ultra-Compact Regime ($\le$350M Parameters)}

\paragraph{SmolLM2-135M}
SmolLM2-135M \cite{allal2025smollm2} represents the extreme edge-deployable regime. Despite its compact size, the model is trained on approximately two trillion tokens using a diversified corpus that includes educational data, code repositories, and curated high-quality text sources.

\paragraph{Granite-4.0-350M}
Granite-350M \cite{granite2025} is a lightweight, instruction-tuned model developed by IBM and derived from a base checkpoint via supervised instruction tuning, reinforcement learning, and model merging.

\paragraph{LFM2-350M}
LFM2-350M \cite{liquidai2025lfm2} introduces architectural diversity into the ultra-compact regime. Unlike purely Transformer-based designs, it employs a hybrid Liquid architecture incorporating multiplicative gating mechanisms and short convolutional blocks.

\subsubsection{Lower Mid-Scale Regime ($\approx$ 1B Parameters)}

\paragraph{Llama 3.2 1B Instruct}
Llama 3.2 1B Instruct \cite{grattafiori2024llama} provides a widely adopted dense-Transformer baseline at the 1B scale. Optimized for efficient multilingual instruction following and low-resource deployment, it serves as a reference model for evaluating the reliability of deterministic reasoning near the hypothesized scaling transition.

\paragraph{Granite-4.0-H-1B}
Granite-4.0-H-1B \cite{granite2025} extends the Granite nano family into the billion-parameter regime while retaining its reinforcement-learning-enhanced alignment pipeline.

\paragraph{LFM2.5-1.2B-Instruct}
LFM2.5-1.2B \cite{liquidai2025lfm2} represents an architectural evolution of the hybrid Liquid design. It combines convolutional LIV blocks with grouped-query attention layers and is trained on an extended 28-trillion-token corpus. The model supports long-context reasoning (up to 32K tokens) while maintaining a memory footprint compatible with edge devices.

\subsubsection{Mid-Scale and Upper-Bound Regime (1.5B–7B)}

\paragraph{Qwen2.5-1.5B-Instruct}
Qwen2.5-1.5B \cite{qwen25,qwen2} marks the lower boundary of the empirically observed stability transition. Architecturally, it employs rotary positional embeddings, SwiGLU activations, RMS normalization, and grouped-query attention. The model is optimized for structured output generation, long-context reasoning, and multilingual robustness.

\paragraph{Qwen2.5-3B-Instruct}
The 3B variant increases depth and attention capacity while preserving architectural design. It enables evaluation of whether scaling beyond the transition region yields qualitative reasoning improvements or primarily incremental refinement \cite{qwen25,qwen2}.

\paragraph{Olmo 3 7B Instruct}
Olmo 3 7B Instruct \cite{olmo2025olmo} is a supervised, instruction-tuned dense Transformer trained under an open, transparent training pipeline. It serves as an upper-bound dense baseline within the compact-large regime, providing insight into performance saturation at higher parameter scales.

\paragraph{Qwen2.5-7B-Instruct}
Qwen2.5-7B \cite{qwen25,qwen2} represents the largest model in our evaluation and provides an empirical ceiling for deterministic semantic reasoning under the 6G-Bench workload. Its inclusion allows precise quantification of diminishing marginal returns relative to the 3B regime.

All models were evaluated locally without reliance on external APIs. Each checkpoint was downloaded from Hugging Face and executed in a controlled inference environment to ensure reproducibility, uniform decoding configurations, and consistent hardware conditions. No additional fine-tuning was performed on 6G-Bench tasks; all results reflect zero-shot semantic reasoning under deployment-realistic constraints.

\subsection{Prompt Template and Context Conditioning Protocol}

For models that support chat formatting, the native tokenizer function \texttt{apply\_chat\_template} is used to construct the input prompt. For other models, a structured system--user concatenation is employed.

The system message defines the evaluator role, e.g.,

\begin{quote}
\end{quote}

\begin{rqbox1}
You are an expert 6G network AI agent evaluator. You will answer a multiple-choice question (A/B/C/D) about a UAV mission episode. Use the episode context and the target task definition to choose the best answer. You \emph{must} respond with a single JSON object of the form \texttt{\{"answer": "A"\}}.
\end{rqbox1}

The user message contains:
\begin{itemize}
    \item \texttt{TASK\_ID}, \texttt{TASK\_NAME}, and the formal task definition,
    \item The automatically generated episode summary,
    \item The MCQ question text,
    \item Options A--D, formatted as \texttt{A: ...}, \texttt{B: ...}, etc.,
    \item A strict instruction to respond only with JSON of the form \texttt{\{"answer": "A/B/C/D"\}}.
\end{itemize}

This uniform prompting protocol ensures that differences in performance arise from model capability and scale rather than prompt variation. Therefore, episode--question pairs are constructed by matching \texttt{.episode.json} files with their corresponding \texttt{.mcq.json} files in the dataset directory. Only episodes that have both files are included in the evaluation.

\subsection{Evaluation Metrics and Stability Measures}

Let $\mathcal{T} = \{t_1, \dots, t_M\}$ denote the evaluation task set,
and let $y_i^\star$ be the ground-truth decision for task $t_i$.
For a model with parameter scale $N$, let $\hat{y}_i^{(j)}(N)$ denote
the $j$-th generated response under stochastic decoding.

\paragraph{Deterministic Accuracy (pass@1)}

Under deterministic decoding (e.g., greedy decoding or temperature $=0$),
each task yields a single prediction $\hat{y}_i(N)$.
We define deterministic single-shot accuracy as

\begin{equation}
A_1(N) =
\frac{1}{M}
\sum_{i=1}^{M}
\mathbf{1}\!\left[\hat{y}_i(N) = y_i^\star\right],
\end{equation}

where $\mathbf{1}[\cdot]$ denotes the indicator function.
This metric reflects safety-critical decision reliability under a
single, policy-consistent inference trajectory. {\color{black}
Since $A_1(N)$ is a binomial proportion over $M$ MCQs,
we report 95\% confidence intervals using the normal approximation
$A_1 \pm 1.96 \sqrt{A_1(1-A_1)/M}$.
}

\paragraph{Stochastic Robustness (pass@k)}

Under stochastic decoding, each task produces $k$ independent samples
$\{\hat{y}_i^{(1)}(N), \dots, \hat{y}_i^{(k)}(N)\}$.
We define pass@k as

\begin{equation}
A_k(N) =
\frac{1}{M}
\sum_{i=1}^{M}
\mathbf{1}
\!\left[
\exists j \in \{1,\dots,k\}
\text{ such that }
\hat{y}_i^{(j)}(N) = y_i^\star
\right],
\end{equation}

which quantifies the probability that at least one reasoning trajectory
among $k$ stochastic samples recovers the correct decision.

\paragraph{Reasoning Instability.}

To characterize stochastic inconsistency, we define the instability gap

\begin{equation}
\Delta_k(N) = A_k(N) - A_1(N),
\end{equation}

which measures the extent to which correct reasoning exists but is not
reliably selected under deterministic decoding. {\color{black}
Confidence intervals for $\Delta_k$ are conservatively estimated by
summing the variances of $A_k$ and $A_1$, yielding slightly wider but
statistically safe uncertainty bounds.
}

All evaluations employ a standardized task-conditioned prompting protocol
with robust answer extraction to ensure comparability across models.

\subsubsection{Log-Linear Scaling Model}

Let $N$ denote the number of model parameters (in billions), and define
$A(N) := A_1(N)$ as deterministic accuracy.
To characterize scaling behavior, we approximate performance as a
log-linear function of model size:

\begin{equation}
A(N) = \alpha \log(1+N) + \beta + \epsilon(N),
\end{equation}

{\color{black}
where $\log(\cdot)$ denotes the natural logarithm.
The inclusion of $1+N$ stabilizes the transformation in the ultra-compact
regime ($N \ll 1$), where $\log N$ becomes strongly negative.
}

where $\alpha$ captures the marginal improvement in semantic reasoning
per logarithmic increase in parameter count, $\beta$ is a baseline offset,
and $\epsilon(N)$ represents architecture-dependent deviations.

{\color{black}
The coefficients $\alpha$ and $\beta$ are estimated via ordinary least
squares (OLS) over all evaluated models. Residuals reported in
Table~\ref{tab:scaling_analysis} are computed as
$A_1 - (\alpha \log(1+N) + \beta)$.
}

To quantify discrete scaling gains, we define

\begin{equation}
\Delta A_{i \rightarrow j} = A(N_j) - A(N_i).
\end{equation}

{\color{black}
Statistical comparisons between scales (e.g., 1B $\rightarrow$ 1.5B)
are performed using two-proportion z-tests on $A_1$,
with significance assessed at $\alpha=0.05$.
}

Empirically, we observe that

\begin{equation}
\Delta A_{1\mathrm{B} \rightarrow 1.5\mathrm{B}}
\gg
\Delta A_{3\mathrm{B} \rightarrow 7\mathrm{B}},
\end{equation}

indicating non-uniform marginal returns and the presence of distinct
scaling regimes rather than smooth proportional growth.

\subsubsection{Group-Level Scaling Sensitivity}

To quantify scaling elasticity across semantic capability domains,
we define a group-level sensitivity coefficient for each capability group $G_j$:

\begin{equation}
S_{G_j} =
\frac{\bar{A}_{G_j}(N_{\max}) - \bar{A}_{G_j}(N_{\min})}
{\log(N_{\max}) - \log(N_{\min})},
\end{equation}

where $\bar{A}_{G_j}(N)$ denotes the mean deterministic accuracy
(pass@1) across all tasks in group $G_j$ at parameter scale $N$.

This coefficient estimates the average improvement in domain-specific
semantic reasoning performance per logarithmic increase in model scale.
By normalizing gains in log-parameter space, $S_{G_j}$ enables
direct comparison of scaling responsiveness across heterogeneous
capability classes.

\begin{table}[t]
\centering
\caption{Experimental configuration and 6G-Bench MCQ dataset statistics.}
\label{tab:exp_config}
\scriptsize
\begin{tabular}{|l|l|}
\hline
\textbf{Component} & \textbf{Configuration / Value} \\ \hline
\multicolumn{2}{|c|}{\textbf{Hardware Configuration}} \\ \hline
GPUs & 8$\times$ NVIDIA A100 (80GB VRAM each) \\ \hline
Framework & PyTorch + Hugging Face Transformers \\ \hline
Precision & bfloat16 (CUDA), float32 (CPU fallback) \\ \hline
Device Placement & \texttt{device\_map="auto"} \\ \hline
Quantization & None \\ \hline
Fine-Tuning & None (zero-shot evaluation) \\ \hline
Model Parameter Range & 0.135B -- 8.34B \\ \hline
Number of Models & 11 \\ \hline
\multicolumn{2}{|c|}{\textbf{Decoding Configuration}} \\ \hline
Deterministic Decoding & Temperature = 0.0, \texttt{do\_sample = False} \\ \hline
Stochastic Decoding & Temperature = 0.7, \texttt{do\_sample = True} \\ \hline
Evaluation Metrics & pass@1 (deterministic), pass@3, pass@5 (stochastic) \\ \hline
Pass@k Values & $k \in \{3,5\}$ \\ \hline
Tasks with Pass@k & 7 (T2, T9, T12, T19, T20, T26, T30) \\ \hline
Max New Tokens & Up to 10{,}000 (bounded by context window) \\ \hline
Seed Control & SHA-256 deterministic seed derivation \\ \hline
\multicolumn{2}{|c|}{\textbf{Inference Profiling Configuration}} \\ \hline
Queries per Model & $R = 20$ runs (single-query profiling) \\ \hline
Prompt Length & 2049 input tokens \\ \hline
Generated Tokens & 8 output tokens (deterministic decoding) \\ \hline
FLOPs Estimation & Forward-pass FLOPs/query (TF) per model \\ \hline
\multicolumn{2}{|c|}{\textbf{6G-Bench MCQ Dataset Statistics}} \\ \hline
Number of Tasks & 30 (T1--T30) \\ \hline
Capability Groups & 5 (G1--G5) \\ \hline
Total Episodes & 488 \\ \hline
Total MCQ Questions & 3{,}722 \\ \hline
Avg. MCQ per Episode & 7.63 \\ \hline
CI Estimation & 95\% binomial CI over $M = 3{,}722$ MCQs \\ \hline
Evaluation Setting & Strict JSON-constrained MCQ \\ \hline
\end{tabular}
\end{table}
\subsubsection{Edge-Oriented Efficiency Metric}

While the preceding metrics characterize semantic reliability and
scaling behavior, practical deployment in AI-native 6G systems
also requires joint consideration of inference latency and memory
footprint under edge constraints.

Let $L(N)$ denote the mean single-query inference latency (ms),
and let $M(N)$ denote the mean peak VRAM usage (GB) for a
model of scale $N$. To quantify semantic reliability per unit
edge resource, we define the \emph{Edge Score}:

\begin{equation}
\mathrm{ES}(N) =
\frac{A_1(N)}{L(N) \cdot M(N)}.
\end{equation}

This metric jointly captures:
\begin{itemize}
    \item deterministic semantic accuracy ($A_1$),
    \item time-to-decision ($L$),
    \item and memory footprint ($M$).
\end{itemize}

Higher $\mathrm{ES}(N)$ indicates greater semantic reliability per unit latency and memory cost, making it directly interpretable as an edge deployment efficiency measure. Unlike purely accuracy-based metrics, the Edge Score explicitly penalizes models that achieve marginal capability gains at the cost of disproportionately higher inference latency or memory overhead. It therefore provides a deployment-aware efficiency frontier that complements the scaling-law analysis. {\color{black}
The Edge Score is intended as a first-order deployment indicator that combines semantic reliability with latency and memory constraints.
}

\section{Empirical Scaling Analysis}
\label{sec:Scaling}

Table~\ref{tab:exp_config} summarizes the complete experimental configuration and dataset statistics used in this study. All evaluations were conducted on a multi-GPU server equipped with eight NVIDIA A100 GPUs (8$\times$80GB VRAM), enabling stable zero-shot inference for models up to 7B parameters without quantization or parameter offloading. Deterministic (pass@1) and stochastic (pass@k) decoding regimes were implemented under strictly controlled seed management to ensure reproducibility. The evaluation corpus comprises 488 episodes and 3,722 multiple-choice questions spanning 30 capability-aligned tasks in 6G-Bench, yielding an average of 7.63 questions per episode. Seven tasks were additionally evaluated under pass@3 and pass@5 to quantify robustness to stochastic reasoning.

\subsection{Global Scaling Trends: Deterministic and Stochastic Performance}

\begin{table*}[t]
\centering
\caption{Scaling analysis of small language models on 6G-Bench.}
\label{tab:scaling_analysis}
\scriptsize
\begin{tabular}{lcccccccccc}
\toprule
Model & $N$ (B) & $\log(1+N)$ & $A_1$ 
& {\color{black}95\% CI ($A_1$)} 
& $A_3$ & $A_5$ & $\Delta_5$ 
& {\color{black}95\% CI ($\Delta_5$)} 
& $\eta(N)$ 
& {\color{black}Residual} \\
\midrule
SmolLM2-135M & 0.135 & 0.127 & 0.224 
& {\color{black}[0.210, 0.238]} 
& 0.440 & 0.589 & 0.365 
& {\color{black}[0.344, 0.386]} 
& \underline{\textbf{1.764}} 
& {\color{black}-0.026} \\

Granite-350M & 0.35 & 0.300 & 0.372 
& {\color{black}[0.356, 0.388]} 
& 0.606 & 0.755 & 0.383 
& {\color{black}[0.362, 0.404]} 
& 1.240 
& {\color{black}+0.015} \\

LFM2-350M & 0.35 & 0.300 & 0.358 
& {\color{black}[0.343, 0.374]} 
& 0.574 & 0.714 & 0.356 
& {\color{black}[0.334, 0.378]} 
& 1.193 
& {\color{black}+0.001} \\

Llama-3.2-1B & 1.0 & 0.693 & 0.373 
& {\color{black}[0.357, 0.389]} 
& 0.603 & 0.729 & 0.356 
& {\color{black}[0.334, 0.378]} 
& 0.538 
& {\color{black}-0.105} \\

Granite-1B & 1.0 & 0.693 & 0.559 
& {\color{black}[0.543, 0.575]} 
& 0.704 & \underline{\textbf{0.776}} & 0.217 
& {\color{black}[0.196, 0.238]} 
& 0.806 
& {\color{black}\underline{\textbf{+0.081}}} \\

LFM2.5-1.2B & 1.2 & 0.788 & 0.530 
& {\color{black}[0.514, 0.546]} 
& 0.606 & 0.676 & 0.146 
& {\color{black}[0.125, 0.167]} 
& 0.673 
& {\color{black}+0.009} \\

Qwen-1.5B & 1.5 & 0.916 & 0.531 
& {\color{black}[0.515, 0.547]} 
& 0.627 & 0.669 & 0.138 
& {\color{black}[0.117, 0.159]} 
& 0.580 
& {\color{black}+0.001} \\

Qwen-3B & 3.0 & 1.386 & 0.643 
& {\color{black}[0.628, 0.658]} 
& 0.678 & 0.694 & 0.051 
& {\color{black}[0.035, 0.067]} 
& 0.464 
& {\color{black}+0.006} \\

Olmo-7B & 7.0 & 2.079 & 0.652 
& {\color{black}[0.637, 0.667]} 
& 0.703 & 0.724 & 0.072 
& {\color{black}[0.055, 0.089]} 
& 0.314 
& {\color{black}-0.065} \\

Qwen-7B & 7.0 & 2.079 & \underline{\textbf{0.707}} 
& {\color{black}[0.692, 0.722]} 
& \underline{\textbf{0.730}} & 0.738 & \underline{\textbf{0.031}} 
& {\color{black}[0.018, 0.044]} 
& 0.340 
& {\color{black}-0.010} \\

\bottomrule
\end{tabular}\\
\textbf{Notation:} $N$ denotes the number of model parameters measured in billions (B). 
$A_1$ represents deterministic single-shot accuracy (pass@1). 
{\color{black}95\% confidence intervals for $A_1$ are computed using a binomial approximation over $M=3{,}722$ MCQs.} 
$A_3$ and $A_5$ represent stochastic success probabilities over three and five samples (pass@3 and pass@5), respectively. 
$\Delta_5 = A_5 - A_1$ denotes the reasoning instability gap. 
{\color{black}Confidence intervals for $\Delta_5$ are conservatively estimated via variance summation of $A_1$ and $A_5$.} 
$\eta(N) = A_1 / \log(1+N)$ denotes log-scale reasoning efficiency. 
{\color{black}Residual denotes $A_1 - (\alpha \log N + \beta)$ with fitted coefficients $\alpha=0.115$, $\beta=0.478$.}
\end{table*}

{\color{black}
Table~\ref{tab:scaling_analysis} shows that deterministic accuracy increases from 0.224 at 135M to 0.707 at 7B parameters, but the gains are concentrated in the 1--1.5B range, where a statistically significant jump ($z\approx13.9$, $p\ll0.001$) indicates a clear empirical scaling transition rather than smooth proportional growth. Concurrently, the instability gap contracts from 0.365 to 0.031, reflecting convergence between stochastic exploration and deterministic inference and suggesting improved reasoning consistency for safety-critical network control. Beyond 3B parameters, scaling yields diminishing returns, primarily improving robustness margins rather than fundamentally altering reasoning behavior. Residual analysis of the fitted log-linear scaling law further reveals architecture-dependent deviations, with models such as Granite-1B and Qwen variants outperforming parameter-count expectations, while Llama-3.2-1B and Olmo-7B underperform, indicating that alignment strategies and training data contribute meaningful second-order effects beyond model size. Accordingly, the observed transition near 1--1.5B parameters should be interpreted as an empirical scaling phenomenon rather than a theoretically established phase boundary.
}

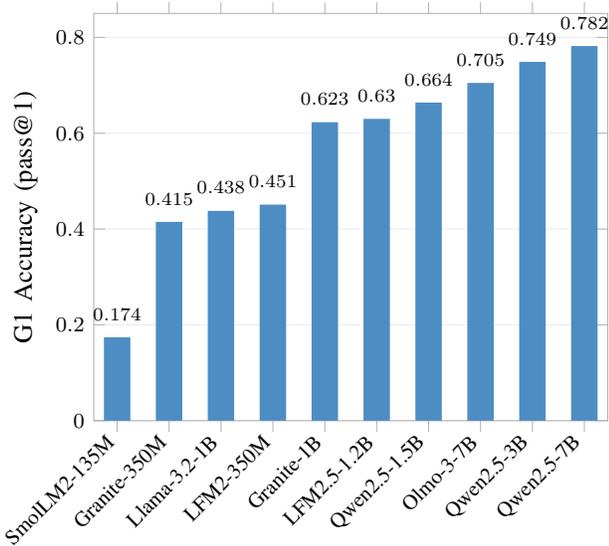
\begin{figure}[t]
\centering
\begin{tikzpicture}
\begin{axis}[
    ybar,
    width=1\linewidth,
    height=5.4cm,
    ymin=0,
    ymax=0.85,
    ylabel={G1 Accuracy (pass@1)},
    symbolic x coords={
        SmolLM2-135M,
        Granite-350M,
        Llama-3.2-1B,
        LFM2-350M,
        Granite-1B,
        LFM2.5-1.2B,
        Qwen2.5-1.5B,
        Olmo-3-7B,
        Qwen2.5-3B,
        Qwen2.5-7B
    },
    xtick=data,
    x tick label style={
        rotate=45,
        anchor=east,
        font=\footnotesize
    },
    yticklabel style={font=\footnotesize},
    bar width=10pt,
    enlarge x limits=0.05,
    ymajorgrids,
    grid style={gray!15},
    axis line style={gray!60},
    tick style={gray!60},
    nodes near coords,
    every node near coord/.append style={
        font=\scriptsize,
        /pgf/number format/fixed,
        /pgf/number format/precision=3,
        yshift=3pt
    }
]

\addplot[
    fill=GOne!80,
    draw=none
] coordinates {
    (SmolLM2-135M,0.174)
    (Granite-350M,0.415)
    (Llama-3.2-1B,0.438)
    (LFM2-350M,0.451)
    (Granite-1B,0.623)
    (LFM2.5-1.2B,0.630)
    (Qwen2.5-1.5B,0.664)
    (Olmo-3-7B,0.705)
    (Qwen2.5-3B,0.749)
    (Qwen2.5-7B,0.782)
};

\end{axis}
\end{tikzpicture}
\caption{Group G1 (Intent \& Policy Reasoning): pass@1 accuracy across model scales.}
\label{fig:g1_intent_policy}
\end{figure}

\begin{figure}[t]
\centering
\begin{tikzpicture}
\begin{axis}[
    ybar,
    width=1\linewidth,
    height=5.4cm,
    ymin=0,
    ymax=0.75,
    ylabel={G2 Accuracy (pass@1)},
    symbolic x coords={
        SmolLM2-135M,
        Granite-350M,
        LFM2-350M,
        Llama-3.2-1B,
        Qwen2.5-1.5B,
        LFM2.5-1.2B,
        Granite-1B,
        Qwen2.5-3B,
        Olmo-3-7B,
        Qwen2.5-7B
    },
    xtick=data,
    x tick label style={
        rotate=45,
        anchor=east,
        font=\footnotesize
    },
    yticklabel style={font=\footnotesize},
    bar width=10pt,
    enlarge x limits=0.05,
    ymajorgrids,
    grid style={gray!15},
    axis line style={gray!60},
    tick style={gray!60},
    nodes near coords,
    every node near coord/.append style={
        font=\scriptsize,
        /pgf/number format/fixed,
        /pgf/number format/precision=3,
        yshift=3pt
    }
]

\addplot[
    fill=GTwo!80,
    draw=none
] coordinates {
    (SmolLM2-135M,0.250)
    (Granite-350M,0.372)
    (LFM2-350M,0.396)
    (Llama-3.2-1B,0.409)
    (Qwen2.5-1.5B,0.532)
    (LFM2.5-1.2B,0.537)
    (Granite-1B,0.560)
    (Qwen2.5-3B,0.609)
    (Olmo-3-7B,0.672)
    (Qwen2.5-7B,0.695)
};

\end{axis}
\end{tikzpicture}
\caption{Group G2 (Network Slicing \& Resource Management): pass@1 accuracy across model scales.}
\label{fig:g2_slicing}
\end{figure}
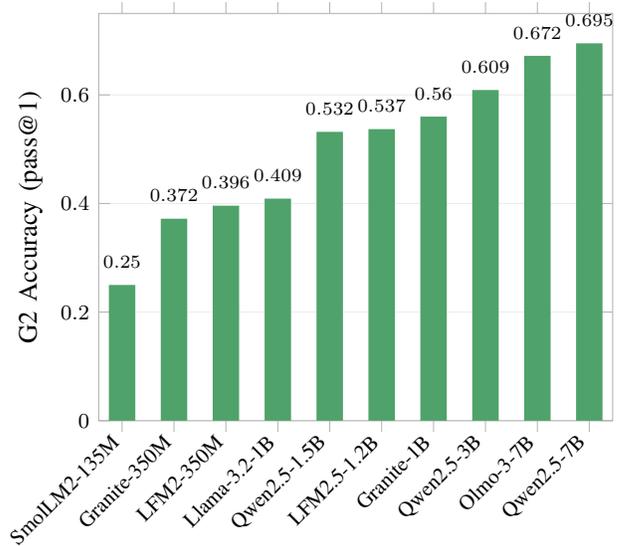

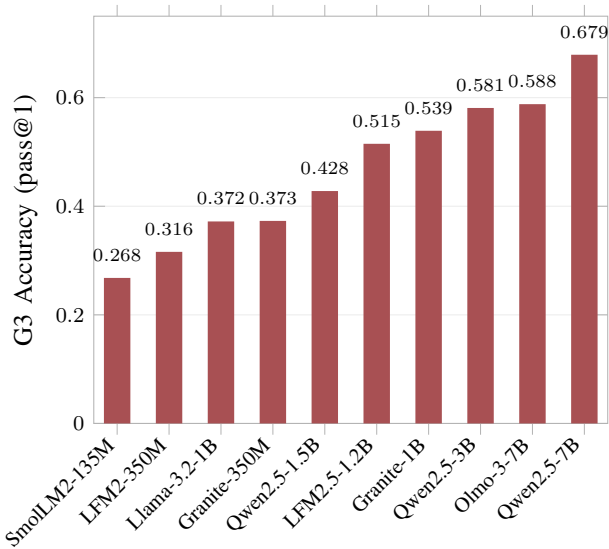
\begin{figure}[t]
\centering
\begin{tikzpicture}
\begin{axis}[
    ybar,
    width=1\linewidth,
    height=5.4cm,
    ymin=0,
    ymax=0.75,
    ylabel={G3 Accuracy (pass@1)},
    symbolic x coords={
        SmolLM2-135M,
        LFM2-350M,
        Llama-3.2-1B,
        Granite-350M,
        Qwen2.5-1.5B,
        LFM2.5-1.2B,
        Granite-1B,
        Qwen2.5-3B,
        Olmo-3-7B,
        Qwen2.5-7B
    },
    xtick=data,
    x tick label style={
        rotate=45,
        anchor=east,
        font=\footnotesize
    },
    yticklabel style={font=\footnotesize},
    bar width=10pt,
    enlarge x limits=0.05,
    ymajorgrids,
    grid style={gray!15},
    axis line style={gray!60},
    tick style={gray!60},
    nodes near coords,
    every node near coord/.append style={
        font=\scriptsize,
        /pgf/number format/fixed,
        /pgf/number format/precision=3,
        yshift=3pt
    }
]

\addplot[
    fill=GThree!80,
    draw=none
] coordinates {
    (SmolLM2-135M,0.268)
    (LFM2-350M,0.316)
    (Llama-3.2-1B,0.372)
    (Granite-350M,0.373)
    (Qwen2.5-1.5B,0.428)
    (LFM2.5-1.2B,0.515)
    (Granite-1B,0.539)
    (Qwen2.5-3B,0.581)
    (Olmo-3-7B,0.588)
    (Qwen2.5-7B,0.679)
};

\end{axis}
\end{tikzpicture}
\caption{Group G3 (Trust, Security \& SLA Awareness): pass@1 accuracy across model scales.}
\label{fig:g3_trust}
\end{figure}

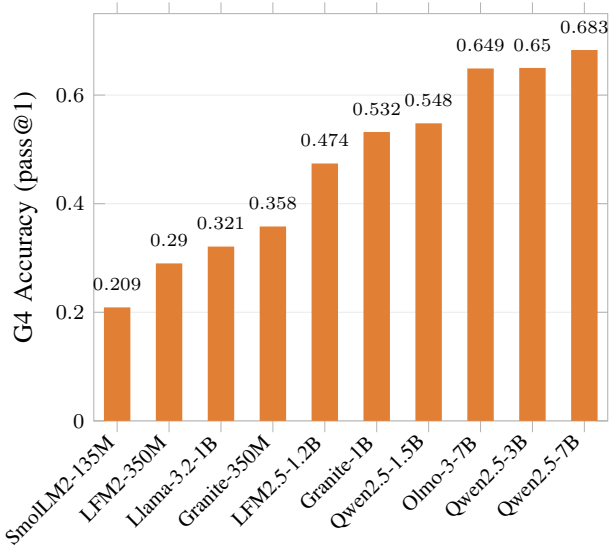
\begin{figure}[t]
\centering
\begin{tikzpicture}
\begin{axis}[
    ybar,
    width=1\linewidth,
    height=5.4cm,
    ymin=0,
    ymax=0.75,
    ylabel={G4 Accuracy (pass@1)},
    symbolic x coords={
        SmolLM2-135M,
        LFM2-350M,
        Llama-3.2-1B,
        Granite-350M,
        LFM2.5-1.2B,
        Granite-1B,
        Qwen2.5-1.5B,
        Olmo-3-7B,
        Qwen2.5-3B,
        Qwen2.5-7B
    },
    xtick=data,
    x tick label style={
        rotate=45,
        anchor=east,
        font=\footnotesize
    },
    yticklabel style={font=\footnotesize},
    bar width=10pt,
    enlarge x limits=0.05,
    ymajorgrids,
    grid style={gray!15},
    axis line style={gray!60},
    tick style={gray!60},
    nodes near coords,
    every node near coord/.append style={
        font=\scriptsize,
        /pgf/number format/fixed,
        /pgf/number format/precision=3,
        yshift=3pt
    }
]

\addplot[
    fill=GFive!80,
    draw=none
] coordinates {
    (SmolLM2-135M,0.209)
    (LFM2-350M,0.290)
    (Llama-3.2-1B,0.321)
    (Granite-350M,0.358)
    (LFM2.5-1.2B,0.474)
    (Granite-1B,0.532)
    (Qwen2.5-1.5B,0.548)
    (Olmo-3-7B,0.649)
    (Qwen2.5-3B,0.650)
    (Qwen2.5-7B,0.683)
};

\end{axis}
\end{tikzpicture}
\caption{Group G4 (AI-Native Networking \& Agentic Control): pass@1 accuracy across model scales.}
\label{fig:g4_agentic}
\end{figure}

\begin{figure}[t]
\centering
\begin{tikzpicture}
\begin{axis}[
    ybar,
    width=1\linewidth,
    height=5.4cm,
    ymin=0,
    ymax=0.75,
    ylabel={G5 Accuracy (pass@1)},
    symbolic x coords={
        SmolLM2-135M,
        Llama-3.2-1B,
        Granite-350M,
        LFM2-350M,
        LFM2.5-1.2B,
        Qwen2.5-1.5B,
        Granite-1B,
        Olmo-3-7B,
        Qwen2.5-3B,
        Qwen2.5-7B
    },
    xtick=data,
    x tick label style={
        rotate=45,
        anchor=east,
        font=\footnotesize
    },
    yticklabel style={font=\footnotesize},
    bar width=10pt,
    enlarge x limits=0.05,
    ymajorgrids,
    grid style={gray!15},
    axis line style={gray!60},
    tick style={gray!60},
    nodes near coords,
    every node near coord/.append style={
        font=\scriptsize,
        /pgf/number format/fixed,
        /pgf/number format/precision=3,
        yshift=3pt
    }
]

\addplot[
    fill=GFour!80,
    draw=none
] coordinates {
    (SmolLM2-135M,0.181)
    (Llama-3.2-1B,0.301)
    (Granite-350M,0.331)
    (LFM2-350M,0.342)
    (LFM2.5-1.2B,0.485)
    (Qwen2.5-1.5B,0.501)
    (Granite-1B,0.541)
    (Olmo-3-7B,0.634)
    (Qwen2.5-3B,0.651)
    (Qwen2.5-7B,0.699)
};

\end{axis}
\end{tikzpicture}
\caption{Group G5 (Distributed Intelligence \& Emerging 6G Use Cases): pass@1 accuracy across model scales.}
\label{fig:g5_distributed}
\end{figure}
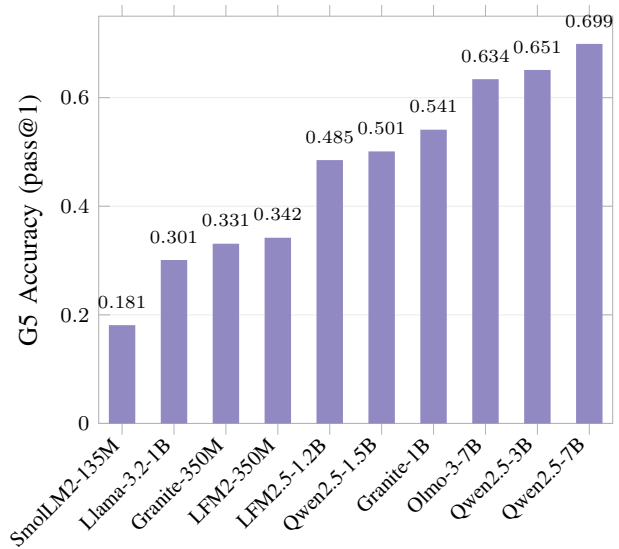

\begin{table*}[t]
\centering
\caption{Group-level scaling sensitivity $S_{G_j}$ between the smallest (SmolLM2-135M, $N_{\min}=0.135$B) and largest (Qwen2.5-7B, $N_{\max}=7$B) models. $\bar{A}_{G_j}(N)$ denotes mean pass@1 over all tasks in group $G_j$.}
\label{tab:group_sensitivity}
\scriptsize
\begin{tabular}{lccc}
\toprule
Group $G_j$ & $\bar{A}_{G_j}(N_{\min})$ & $\bar{A}_{G_j}(N_{\max})$ & $S_{G_j}$ \\
\midrule
G1 -- Intent \& Policy Reasoning                      & 0.174 & 0.782 & 0.154 \\
G2 -- Network Slicing \& Resource Management          & 0.250 & 0.695 & 0.113 \\
G3 -- Trust, Security \& SLA Awareness                & 0.268 & 0.679 & 0.104 \\
G4 -- AI-Native Networking \& Agentic Control         & 0.209 & 0.683 & 0.120 \\
G5 -- Distributed Intelligence \& Emerging 6G Use Cases & 0.181 & 0.699 & 0.131 \\
\bottomrule
\end{tabular}
\end{table*}

\subsection{Capability-Level Results}

{\color{black}Complete per-task results (T1–T30), including model responses, task-level pass@1 accuracies, and full evaluation logs for all models, are publicly available in our accompanying GitHub repository. Due to space constraints, we report aggregated and group-level analyses in our paper.}

\subsubsection{G1 -- Intent \& Policy Reasoning}

Figure~\ref{fig:g1_intent_policy} presents the pass@1 accuracy for Group~G1 (Intent \& Policy Reasoning), capturing models’ ability to perform multi-step intent interpretation, constraint alignment, and policy-consistent decision synthesis. A clear capacity-driven hierarchy emerges: Qwen2.5-7B achieves the highest accuracy (0.782), followed by Qwen2.5-3B (0.749) and Olmo-3-7B (0.705), indicating that both parameter scale and architectural efficiency contribute to performance in intent-centric reasoning. Notably, the 3B Qwen variant approaches 7B-level performance, suggesting strong parameter utilization efficiency. In contrast, sub-billion-parameter models exhibit pronounced degradation, with Llama-3.2-1B (0.438) and SmolLM2-135M (0.174) demonstrating substantial reasoning collapse under policy consistency constraints. The nearly fourfold improvement from 135M to 7B underscores the strong dependence of intent alignment and uncertainty-aware reasoning on representational capacity, indicating that policy-consistent semantic abstraction is intrinsically scale-sensitive.

\subsubsection{G2 -- Network Slicing \& Resource Management}

Figure~\ref{fig:g2_slicing} presents the results for Group~G2 (Network Slicing \& Resource Management), reflecting structured resource allocation, constraint satisfaction, and cross-layer optimization reasoning. While performance scales monotonically with parameter count, the gradient is less steep than in G1. Qwen2.5-7B leads (0.695), closely followed by Olmo-3-7B (0.672), suggesting that structured allocation reasoning benefits from scale but exhibits earlier saturation dynamics. The gap between 7B and 3B models is comparatively narrow, implying that slicing-related reasoning relies partially on structured constraint modeling rather than exclusively on deep latent abstraction. Sub-1B models retain moderate capability (0.40--0.56 range), indicating that these tasks remain partially tractable under constrained parameterization. Overall, G2 demonstrates steady yet comparatively moderate scaling behavior characteristic of semi-structured optimization domains.

\subsubsection{G3 -- Trust, Security \& SLA Awareness}

Figure~\ref{fig:g3_trust} presents the pass@1 performance for Group~G3 (Trust, Security \& SLA Awareness), which emphasizes compliance verification, risk assessment, and service-level reasoning under adversarial or contractual constraints. The highest accuracy is obtained by Qwen2.5-7B (0.679), yet the relative improvement over mid-scale models such as Qwen2.5-3B (0.581) is less pronounced than in G1. LiquidAI LFM2.5-1.2B (0.515) performs competitively with several larger counterparts, suggesting that structured policy conditioning and the composition of training data may partially compensate for parameter scale in trust-oriented tasks. Although degradation toward ultra-small models remains evident (SmolLM2-135M at 0.268), the decline is comparatively smoother than in G1. This pattern indicates lower scaling elasticity and suggests that trust and SLA reasoning may saturate earlier with scale, potentially reflecting a stronger dependence on structured rule internalization than on increasingly deep generative abstraction.

\subsubsection{G4 -- AI-Native Networking \& Agentic Control}

Figure~\ref{fig:g4_agentic} presents the results for Group~G4 (AI-Native Networking \& Agentic Control), capturing sequential decision-making, autonomous orchestration, and control-loop reasoning. The top-performing models—Qwen2.5-7B (0.683), Qwen2.5-3B (0.650), and Olmo-3-7B (0.649)—form a tightly clustered high-performance regime, indicating that agentic control benefits from both scale and architectural design. The moderate performance gap between 7B and 3B models suggests diminishing marginal returns at higher capacity, possibly reflecting partial saturation in control-sequence abstraction. However, a pronounced drop occurs below the 1B threshold (e.g., Llama-3.2-1B at 0.321), highlighting the importance of sufficient internal state representation for coherent multi-step orchestration. Performance dispersion across smaller models further suggests sensitivity to instruction-tuning and alignment mechanisms in agentic reasoning contexts.

\subsubsection{G5 -- Distributed Intelligence \& Emerging 6G Use Cases}

Figure~\ref{fig:g5_distributed} presents the pass@1 accuracy for Group~G5 (Distributed Intelligence \& Emerging 6G Use Cases), encompassing federated coordination, Integrated Sensing and Communication (ISAC), and digital twin synchronization. Scaling effects are again pronounced: Qwen2.5-7B achieves 0.699, whereas the smallest model attains only 0.181. The 3B variant (0.651) demonstrates strong parameter efficiency; however, the absolute improvement across scales remains substantial, indicating that distributed reasoning and cross-entity coordination impose significant representational demands. Compared to G2 and G3, where structured constraints partially mitigate scale limitations, G5 tasks appear to require deeper semantic integration across heterogeneous abstractions. This observation aligns with the inherent complexity of emerging 6G scenarios that involve multi-domain coordination and distributed cognition.

\subsubsection{Scaling Sensitivity Across Capability Domains}

Table~\ref{tab:group_sensitivity} presents the scaling sensitivity coefficients $S_{G_j}$ computed between the smallest model (SmolLM2-135M, $N_{\min}=0.135$B) and the largest model (Qwen2.5-7B, $N_{\max}=7$B). The results reveal clear heterogeneity in how semantic capability classes benefit from parameter scaling. Intent \& Policy Reasoning (G1) exhibits the highest sensitivity ($S_{G1}=0.154$), indicating that multi-step intent alignment and consistency under uncertainty improve most strongly with increasing model capacity. Distributed Intelligence \& Emerging 6G Use Cases (G5) also shows pronounced scaling dependence ($S_{G5}=0.131$), reflecting the complexity of reasoning required for federated learning orchestration, Integrated Sensing and Communication (ISAC), and digital twin coordination. In contrast, Trust, Security \& Service-Level Agreement (SLA) Awareness (G3) demonstrates the lowest sensitivity ($S_{G3}=0.104$), suggesting that these tasks rely less on raw parameter scaling and more on structured policy reasoning and alignment mechanisms. Network Slicing \& Resource Management (G2) and AI-Native Networking \& Agentic Control (G4) occupy intermediate positions, exhibiting steady but moderate scaling gains. These results quantitatively confirm the heterogeneous scaling elasticity observed across semantic domains.

\begin{table*}[ht]
\centering
\caption{Single-query inference profile (mean $\pm$ std) for LLMs evaluated on 6G-Bench.}
\label{tab:6gbench_inference_profile}
\scriptsize
\begin{tabular}{lccccc}
\toprule
Model & Params (B) & Peak VRAM (GB) & Latency (ms) & Throughput (tok/s) & FLOPs/query (TF) \\
\midrule
HuggingFaceTB/SmolLM2-135M-Instruct & 0.135 & 0.331 $\pm$ 0.000 & \underline{\textbf{50.4}} $\pm$ 100.8 & 68.7 $\pm$ 15.8 & \underline{\textbf{1.692}} $\pm$ 0.000 \\
LiquidAI/LFM2-350M & 0.354 & \underline{\textbf{0.137}} $\pm$ 0.001 & 136.3 $\pm$ 154.1 & 75.1 $\pm$ 15.3 & 4.375 $\pm$ 0.000 \\
ibm-granite/granite-4.0-h-350m & 0.340 & 16.194 $\pm$ 0.000 & 1814.7 $\pm$ 128.7 & 3.9 $\pm$ 0.2 & 4.074 $\pm$ 0.000 \\
meta-llama/Llama-3.2-1B-Instruct & 1.236 & 2.481 $\pm$ 0.000 & 100.1 $\pm$ 111.2 & \underline{\textbf{89.0}} $\pm$ 18.1 & 13.688 $\pm$ 0.000 \\
LiquidAI/LFM2.5-1.2B-Instruct & 1.170 & 0.266 $\pm$ 0.002 & 152.5 $\pm$ 153.0 & 64.7 $\pm$ 12.9 & 14.444 $\pm$ 0.000 \\
ibm-granite/granite-4.0-h-1b & 1.462 & 20.371 $\pm$ 0.000 & 3161.3 $\pm$ 63.3 & 2.2 $\pm$ 0.0 & 17.495 $\pm$ 0.000 \\
Qwen/Qwen2.5-1.5B-Instruct & 1.544 & 0.449 $\pm$ 0.001 & 209.7 $\pm$ 179.1 & 40.1 $\pm$ 8.9 & 17.654 $\pm$ 0.000 \\
Qwen/Qwen2.5-3B-Instruct & 3.086 & 0.874 $\pm$ 0.000 & 273.9 $\pm$ 181.1 & 29.2 $\pm$ 6.0 & 35.291 $\pm$ 0.000 \\
allenai/Olmo-3-7B-Instruct & 7.298 & 1.336 $\pm$ 0.000 & 386.3 $\pm$ 180.1 & 19.5 $\pm$ 3.2 & 87.357 $\pm$ 0.000 \\
Qwen/Qwen2.5-7B-Instruct & 7.616 & 1.028 $\pm$ 0.000 & 338.7 $\pm$ 202.3 & 23.0 $\pm$ 4.0 & 87.092 $\pm$ 0.000 \\
LiquidAI/LFM2-8B-A1B & 8.340 & 1.275 $\pm$ 0.000 & 552.2 $\pm$ 295.8 & 15.9 $\pm$ 2.9 & 102.931 $\pm$ 0.000 \\
\bottomrule
\end{tabular}\\
\textbf{Notes:} Results are mean $\pm$ standard deviation over $R=20$ runs.
Prompt length: 2049 input tokens; generated tokens: 8.
Measurements collected with deterministic decoding (temperature=0), \texttt{bf16}, no quantization.
All results correspond to single-GPU execution. {\color{black}
The LFM2-8B-A1B model is included only for inference-cost profiling and is not part of the 6G-Bench accuracy and scaling evaluation.
}
\end{table*}

\begin{figure}[t]
\centering
\begin{tikzpicture}
\begin{axis}[
    ybar,
    width=1\linewidth,
    height=5.4cm,
    ymin=0,
    ymax=210,
    ylabel={Edge Score ($\times 10^{4}$)},
    symbolic x coords={
        Granite-1B,
        Granite-350M,
        Olmo-3-7B,
        Llama-3.2-1B,
        Qwen2.5-7B,
        Qwen2.5-3B,
        Qwen2.5-1.5B,
        LFM2.5-1.2B,
        SmolLM2-135M,
        LFM2-350M
    },
    xtick=data,
    x tick label style={
        rotate=45,
        anchor=east,
        font=\footnotesize
    },
    yticklabel style={font=\footnotesize},
    bar width=10pt,
    enlarge x limits=0.05,
    ymajorgrids,
    grid style={gray!15},
    axis line style={gray!60},
    tick style={gray!60},
    nodes near coords,
    every node near coord/.append style={
        font=\scriptsize,
        /pgf/number format/fixed,
        /pgf/number format/precision=1,
        yshift=3pt
    }
]

\addplot[
    fill=EdgeColor!80,
    draw=none
] coordinates {
    (Granite-1B,0.1)
    (Granite-350M,0.1)
    (Olmo-3-7B,12.6)
    (Llama-3.2-1B,15.0)
    (Qwen2.5-7B,20.3)
    (Qwen2.5-3B,26.9)
    (Qwen2.5-1.5B,56.4)
    (LFM2.5-1.2B,130.7)
    (SmolLM2-135M,134.3)
    (LFM2-350M,191.7)
};

\end{axis}
\end{tikzpicture}
\caption{Edge Score ranking of evaluated LLMs on 6G-Bench. Edge Score is defined as $\mathrm{ES} = A_1/(L \cdot M)$, where $A_1$ is deterministic accuracy, $L$ is latency (ms), and $M$ is peak VRAM (GB). Scores are scaled by $10^{4}$; higher values indicate better semantic reliability per unit edge resource.}
\label{fig:edge_score}
\end{figure}
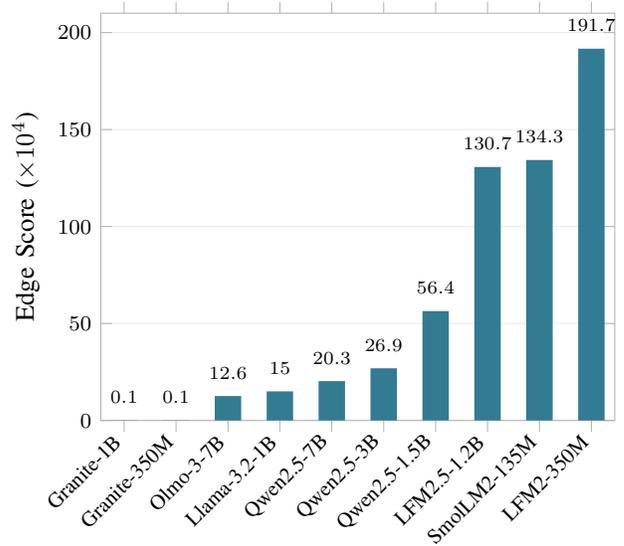

\subsection{Inference Cost and Efficiency Profiling}
\label{subsec:inference_profile}

Table~\ref{tab:6gbench_inference_profile} reports single-query inference profiling across all evaluated LLMs under deterministic decoding. Latency and throughput scale nonlinearly with parameter count: ultra-compact models (135M--350M) achieve sub-150\,ms latency and the lowest memory footprints, while 7B--8B models incur 300--550\,ms latency and substantially higher compute cost. Peak VRAM does not strictly follow parameter size, with certain architectures exhibiting disproportionately large memory footprints, indicating implementation-dependent overheads. Throughput generally declines beyond the 1--3B regime, reflecting diminishing efficiency gains at larger scales. Theoretical forward-pass FLOPs per query increase approximately linearly with model size, ranging from 1.69\,TF (135M) to over 100\,TF (8B). These results highlight a clear trade-off between semantic capability gains and inference cost, suggesting that mid-scale models (1.5--3B) provide a favorable balance between stability and deployability in resource-constrained 6G edge environments.

\subsection{Design Implications for Edge Deployment}

Figure~\ref{fig:edge_score} ranks the evaluated models according to the Edge Score, $\mathrm{ES}=A_1/(L \cdot M)$, which measures deterministic semantic reliability per unit latency and memory footprint. The results reveal a strongly non-monotonic relationship between parameter scale and edge-normalized efficiency. Ultra-compact and hybrid models dominate the frontier: LFM2-350M achieves the highest score (191.7$\times10^{4}$), followed by SmolLM2-135M (134.3) and LFM2.5-1.2B (130.7), indicating that architectural efficiency and low resource consumption can outweigh raw parameter capacity in edge settings. In contrast, larger dense models—including Qwen2.5-7B (20.3) and Olmo-3-7B (12.6)—exhibit substantially lower deployment-normalized efficiency despite higher absolute accuracy, reflecting the compounding cost of latency and VRAM usage. Notably, Qwen2.5-1.5B (56.4) and Qwen2.5-3B (26.9) illustrate the trade-off near the stability transition: while deterministic reliability improves with scale, edge efficiency declines beyond the mid-scale regime. {\color{black}
Overall, the figure provides a comparative view of deployment efficiency under the evaluated hardware configuration. Since profiling was conducted on NVIDIA A100 GPUs, the reported values should not be interpreted as direct performance estimates for operational edge devices.
}

\subsection{Limitations and Reliability Concerns}

{\color{black}
While this study focuses on deterministic scaling behavior, the instability gap ($\Delta_5$) indicates that ultra-compact models may contain correct reasoning trajectories that are not reliably selected by deterministic decoding, posing risks for safety-critical tasks such as feasibility assessment, SLA prediction, and incident response. Even at 3B--7B, residual instability motivates hybrid deployments combining LLMs with verifiers, rule-based safeguards, and constrained decoding, particularly in safety-sensitive domains. Additional limitations include the use of a single prompting protocol, the non-exhaustive coverage of model architectures, and the bounded scope of 6G-Bench, which may not fully represent future multimodal, adversarial, or operationally diverse 6G environments. Furthermore, differences in training data, alignment procedures, and optimization objectives introduce architecture-dependent deviations from ideal log-linear scaling. As 6G-Bench was developed by our research group, benchmark-author bias cannot be completely excluded; however, to support reproducibility and independent validation, all benchmark resources, evaluation scripts, and results have been publicly released at \url{https://github.com/maferrag/6G-Bench}. Despite these limitations, the identified scaling regimes provide a useful foundation for designing capability-aware semantic control layers in AI-native 6G systems.
}

\section{Conclusion}
\label{sec:conclu}

{\color{black}
This paper presents the first systematic scaling study of small and tiny language models (135M--7B parameters) for network-level semantic reasoning in AI-native 6G systems using the 6G-Bench framework. Results reveal a clear stability transition around 1--1.5B parameters, diminishing returns beyond 3B, and a substantial reduction in the instability gap from 0.365 to 0.031, indicating convergence between stochastic exploration and deterministic inference. While intent reasoning and distributed intelligence remain highly scale-sensitive, trust and resource management capabilities saturate earlier. Overall, the findings identify mid-scale models (1.5--3B) as an effective balance between computational efficiency and deterministic reliability, providing practical guidance for deploying LLM-based agents across 6G network hierarchies and highlighting that the key objective is not maximum model size, but the minimum stable capacity required for reliable semantic control in AI-native 6G networks.
}

\bibliographystyle{IEEEtran}
\bibliography{bibliography}

\end{document}